\documentclass{article}

\usepackage[english]{babel}

\usepackage[letterpaper,top=2cm,bottom=2cm,left=3cm,right=3cm,marginparwidth=1.75cm]{geometry}

\usepackage{amsthm}
\usepackage{verbatim,color,amssymb}
\usepackage{amsmath}					
\usepackage{amsthm}					
\usepackage{algorithm,algorithmic}
\usepackage[round]{natbib}
\usepackage{setspace}
\usepackage[mathscr]{euscript}
\usepackage{fancyhdr}
\usepackage{enumitem}
\usepackage{graphicx}
\usepackage{geometry}
\usepackage{lineno}
\usepackage[compact]{titlesec}
\usepackage{listings}
\usepackage{rotating}
\usepackage{booktabs}

\usepackage{mathtools}
\usepackage{subfig}
\usepackage[utf8]{inputenc}
\usepackage[T1]{fontenc}
\usepackage{float}
\usepackage{tikz}
\usetikzlibrary{arrows,chains,backgrounds,fit}
\usepackage{multirow}
\usepackage{lineno}

\newtheorem*{Proof*}{Proof}

\def\bzero{{\mathbf 0}}

\def\var{\hbox{var}}
\def\cov{\hbox{cov}}

\def\Dir{\hbox{Dirch}}
\def\Exp{\hbox{Exp}}

\def\Ga{\hbox{Ga}}

\def\MVN{\hbox{MVN}}

\def\Normal{\hbox{Normal}}

\def\Unif{\hbox{Unif}}

\def\var{\hbox{var}}
\def\cov{\hbox{cov}}

\def\Normal{\hbox{Normal}}

\def\P_25_ICML{{\it Proceedings of the 25th international conference on Machine learning}}

\def\bse{\begin{eqnarray*}}
	\def\ese{\end{eqnarray*}}
\def\be{\begin{eqnarray}}
\def\ee{\end{eqnarray}}
\def\bq{\begin{equation}}
\def\eq{\end{equation}}

\def\trans{^{\rm T}}

\def\b1e{{\mathbf e}}

\def\bb{{\mathbf b}}

\def\bI{{\mathbf I}}

\def\bq{{\mathbf q}}

\def\bX{{\mathbf X}}

\def\bzero{{\mathbf 0}}

\newcommand{\bbeta}{\mbox{\boldmath $\beta$}}

\newcommand{\uB}       {\mbox{\boldmath$B$}}
\newcommand{\ub}       {\mbox{\boldmath$b$}}

\newcommand{\uG}       {\mbox{\boldmath$G$}}

\newcommand{\uM}       {\mbox{\boldmath$M$}}

\newcommand{\uP}       {\mbox{\boldmath$P$}}

\newcommand{\us}       {\mbox{\boldmath$s$}}

\newcommand{\uU}       {\mbox{\boldmath$U$}}

\newcommand{\uW}       {\mbox{\boldmath$W$}}

\newcommand{\uX}       {\mbox{\boldmath$X$}}

\newcommand{\uY}       {\mbox{\boldmath$Y$}}

\newcommand{\uZ}       {\mbox{\boldmath$Z$}}

\newcommand{\ubeta}             {\mbox{\boldmath$\beta$}}

\newcommand{\utheta}            {\mbox{\boldmath$\theta$}}

\newcommand{\uiota}             {\mbox{\boldmath$\uiota$}}

\newcommand{\umu}               {\mbox{\boldmath$\mu$}}
\newcommand{\unu}               {\mbox{\boldmath$\nu$}}

\newcommand{\upi}               {\mbox{\boldmath$\pi$}}

\newcommand{\uphi}              {\mbox{\boldmath$\phi$}}

\newcommand{\uomega}            {\mbox{\boldmath$\omega$}}

\newcommand{\uOmega}            {\mbox{\boldmath$\Omega$}}

\graphicspath{ {/Plots/} }
\usepackage{amsmath}
\usepackage{graphicx}
\usepackage[colorlinks=true, allcolors=blue]{hyperref}

\title{A Bayesian Semi-Parametric Scalar-On-Function Quantile Regression with Measurement Error using the GAL distribution}
\author{Roger S. Zoh, Annie Yu,  Carmen Tekwe}

\begin{document}
\maketitle

\begin{abstract}
Quantile regression provides a consistent approach to investigating the association between covariates and various aspects of the distribution of the response beyond the mean. When the regression covariates are measured with errors, measurement error (ME) adjustment steps are needed for valid inference. This is true for both scalar and functional covariates. Here, we propose extending the Bayesian measurement error and Bayesian quantile regression literature to allow for available covariates prone to potential complex measurement errors. Our approach uses the Generalized Asymmetric Laplace (GAL) distribution as a working likelihood. The family of GAL distribution has recently emerged as a more flexible distribution family in the Bayesian quantile regression modeling compared to their Asymmetric Laplace (AL) counterpart. We then compared and contrasted two approaches in our ME-adjusted steps through a battery of simulation scenarios. Finally, we apply our approach to the analysis of an NHANES dataset 2013-2014 to model quantiles of Body mass index (BMI) as a function of minute-level device-based physical activity in a cohort of an adult 50 years and above..    
\end{abstract}

\section{Introduction}
Quantile regression methodology provides a framework to explore the association between various distributional aspect of the response variable $Y$ and a set of covariates $\bX$, beyond the traditional mean regression. The literature on quantile regression methodology continues to grow with application covering a wide array of fields economics \cite{fitzenberger2001economic}; education \cite{martins2004does}; medicine \cite{wei2006quantile,hong2019quantile}, just to cite a few. \citep{koenker2017quantile, koenker2017handbook}  provide a nice summaries of the methodological advances made in quantile regressions.

Estimation of quantile regression parameters is overwhelmingly addressed from a frequentist view point \cite{koenker2017quantile}. This is in part due to the non-parametric feature of quantile regression, in which no parametric distribution is explicitly posited. And estimation is based on minimization of an objective function (check/pintball function). 

In this paper, we will primarily focus on a Bayesian approach. Bayesian approaches allow propagation of uncertainty in all aspect of the model, taking into account all model uncertainty in the inference.  Bayesian quantile regression remain a very active area of research.  Various approaches have been considered including Bayesian non-parametric approach based on Dirichlet process mixture \cite{gelfand2002computational}; semi-parametric approach \cite{reich2013bayesian,lancaster2010bayesian}; fully parametric with the family of Asymmetric Laplace Distribution (ALD) emerging as a candidate likelihood in Bayesian quantile regression \cite{yu2001bayesian,kozumi2011gibbs}. Concerns, however, have arise about the flexibility of the ALD family \cite{yan2017new} and the validity of posterior inference based on ALD likelihood \cite{yang2016posterior}.  Namely,  for example,  the distribution is always symmetric for the median.and the  mode of the distribution is determined by the location parameter of the distribution. These rather constraining behavior have casted serious doubt of the flexible of ALD as a viable likelihood in quantile regression.  Instead, Generalized Asymmetric Laplace (GAL) distribution has emmerged as a possible likelihood in a Bayesian quantile regression \cite{yan2017new,rahman2019flexible,kobayashi2021flexible}.  Although the use of GAL is growing as a working liklihood for Bayesian quantile regression, it has not yet been using in a case of quantile regression with functional covariates measured with potential complex measurement error.

With our paper, we set out to fill in the gap that exists in the Bayesian estimation of a scalar-on-function quantile regression with measurement error. To the best of our knowledge, no method currently exist in the Bayesian literature to estimate the effect of a functional covariate on various aspect of distribution of the outcome, especially when these functional covariates are mis-measured with potential complex heteroscedatic error. To add more flexible to our model,   our approach uses a truncated Dirichlet process mixture (tDPM) of GAL as a likelihood for the responses. Subsequently, we consider two approaches to correct for measurement errors: a joint model based approach, similar to our previous work \cite{zoh2022fully}, and a fast regression calibration approach motivated by the work of \cite{cui2022fast}. The paper is organized as follows. In section~\ref{sec:model} we briefly introduce the motivating problem and our model and estimation; section~\ref{sec:simul} discuss the simulation settings and the results. We then apply our proposed model to the analysis of the NHANSES 2013-2014 data to estimate the effect of device measure PA on the distribution of BMI in section~\ref{sec:application} and provide concluding remarks in section~\ref{sec:concl}.

\section{Motivating example}
The model proposed in this paper is motivated from the need to assess the impact of  monitor based physical activity at various quantile of BMI for a cohort of  US adult in the NHANSES data set. The data we consider is the publicly available nutritional health and examination survey (NHANES) 2013-2014. NHANES is a survey that examines a nationally representative sample of around 5000 each year. The survey includes demographic questionnaires, examination questionnaires, and laboratory tests conducted by the center for Disease Control and Prevention(CDC). For our analysis, we extract BMI and other demographic information from the demographics data. In addition to demographics information, we also obtained physical activity monitor(PAM) data from the examination dataset. NHANES introduced PAM data since 2011. Study participant were asked to wear a monitor that measure acceleration in 3 axes (x-,y-,z-) at 80Hz (every 1/80 second) for 24hrs for at least 7 days. Data are then downloaded from the devices when they are returned at the end of the study. The devices are checked for malfunction. Then the data are then downloaded and process for analysis.         

\section{Model Specification}   \label{sec:model}
Suppose the following data is obtain from individual $i = 1, \cdots, n$ $\mathcal{D}_{i} = (Y_{i}, \uW_i)$, where $Y_{i} \in \mathbb{R}$, $\uW_i \in \mathbb{R}^{T}$ and $T$ is a number distinct time points at which we observe $\uW_i$. Similar to the model propose by \cite{Tek2022Biost}, the scalar-on-function $\tau$th-quantile regression with measurement error is obtained as
\be
Q_{\tau}(Y_{i}|\uX,\uZ) &=& \widetilde{\beta}_{0,\tau} + \uZ_i\trans\widetilde{\utheta}_{\tau} + \int_{\mathcal{T}} \uX_{i}(t)\ubeta_{1,\tau}(t)dt \label{eq1} \\
W_{ij}(t) &=& X_{i}(t) + U_{ij}(t),\;j=1,2, \cdots,J \label{eq2}
\ee
$\uZ_{i}$ is the vector of error free covariates of length $p$ and $\widetilde{\utheta}$ their associated coefficients; $\uX_i = \{X_{i}(t_1), \cdots, X_{i}(t_{T})\}$ is the functional covariate and $\ubeta_{1,\tau}(t)$, their (functional) effects on the $\tau$th quantile of $Y$. Unfortunately, $\uX_i$ is not truly observed, but its (unbiased) proxy $\uW_{ij} = \{W_{ij}(t_1), \cdots, W_{ij}(t_{T_i}) \}$ is. We assume that $J \geq 2$ denotes the number of replicated measures of $\uW$. Equation~\ref{eq2} describes the measurement error model and the error term $U_{ij}(t)$ is assumed to come from a Gaussian process with mean zero $E\{U_{ij}(t) \} = 0$ and potentially complex/arbitrary covariance function. We only impose that $\cov\{U_{ij_1}(t_j), U_{ij_2}(t_l)\} < \infty$, for any two distinct time points $t_j \neq t_l$. 
Estimation of the models in Equation~\ref{eq1} and \ref{eq2} is difficult for many reasons including the high (infinite) dimension of the parameters (functional parameters $\ubeta_{1,\tau}(t)$) and lack of replicate of the proxy $\uW_{ij}$(i.e, $J =1 $). For the case where $J =1$ (no replicate for $\uW_i$), and \cite{zoh2022fully} and \cite{Tek2022Biost} proposed an approach to estimating model parameters in Eqs~\ref{eq1} and \ref{eq2} using an instrumental variable(IV). A similar approach can be easily be implemented in our Bayesian model with no added complexity. 
We approach parameters space reduction using a basis expansion approach. Namely, if we denote by $\uB_{i,K_n} = (\bb_{i,1},\cdots, \bb_{i,K_n}) \in \mathbb{R}^{T_i \times K_n}$, $\ub_{k}$ is the $k$th basis function $(k=1,\cdots, K_n)$; $T_i$ is the number of distinct time at which the functional data $\uW_i$ is observed for individual $i$ and $K_n$ the number of basis function. The parameter reduced form of Eqs~\ref{eq1} and \ref{eq2} then becomes 
\be
Y_{i}|\uX_i, \uZ_i &=& \beta_{0} + \uZ_i\trans\ubeta_z + \sum^{K_n}_{k=1} \phi_{k}X_{i,k} + \epsilon_{i} \label{eq4} \\
W_{i,k,j} &=& X_{i,k} + U_{i,k,j}, \label{eq5}
\ee
with $i=1,\cdots, n$; $j = 1, \cdots, J$, and $k = 1, \cdots, K_n$ where $X_{i,k} = \int \bb_{i,k}(t)X_{i}(t)dt$; \textcolor{red}{$\phi_{k} = \int \bb_{i,k}(t)\beta_{1}(t)dt$} ; $\epsilon_i$ has a density function $f_{\epsilon}(.)$ such that $\int^{0}_{-\infty} f_{\epsilon}(\epsilon) = \tau$;  $W_{i,k} = \int \bb_{i,k}(t)W_{i}(t)dt$; $U_{i,k} = \int \bb_{i,k}(t)U_{i}(t)dt$. In our Bayesian setting, the quantile regression model specification is concluded with the specification of a flexible distribution for the error term $\epsilon_i$.  Our full Bayesian approach will proceed with models in Equations~\ref{eq4} and \ref{eq5}. 

\subsection{Generalized Asymmetric Laplace (GAL)}
In light of the issues enumerated with ALD in Bayesian quantile regression, other alternatives for the error distribution have emerged. Recently, \cite{yan2017new} proposed the GAl distribution as an alternative for the error term in the Bayesian quantile setting. GAL is obtained as a mixture distribution of a skew-normal distribution and a mixing truncated normal distribution on the positive real line \citep{azzalini2012some}. Namely, if $\epsilon$ has a GAL distribution, then $\epsilon$ admits the following mixture representation
\be
\epsilon = \alpha s + A\nu + u\sqrt{\sigma B\nu},&\;&\text{where}\; s \sim \Normal^{+}(0, \sigma^2),\; \nu \sim \text{Exp}(\sigma),\; u \sim \Normal(0, 1),   
\ee
where $\alpha \in \mathbb{R}$ is the skewness parameter; $\Normal(a,b)$ denotes the normal distribution with mean $a$ and variance $b$, $\Exp(.|a)$ denote the exponential with rate $a$, and $\Normal^{+}(.|0,1)$ denote the standard normal distribution truncated on the positive real line. 
GAL provides more flexibility than ALD in modeling quantiles. Namely, if $\alpha \in \mathbf{R}$, $\mu \in \mathcal{R}$, $\sigma >0$, then $\epsilon$ has a GAL if its density can be obtained as 
\be 
f_{GAL}(\epsilon|\tau, \alpha, \sigma) &=& \frac{2\tau(1-\tau)}{\sigma}\left(  \left[ \Phi(\epsilon^{*}/\alpha - \alpha p_{\alpha-}) - \Phi(- \alpha p_{\alpha-})\right]\exp\left\{ - \epsilon^{*} p_{\alpha-}  + (- \alpha p_{\alpha-})^{2}/2 \right\} \right. \nonumber \\ 
&&\left. \times I(\epsilon^{*}/\alpha > 0) +  \Phi\left( - \alpha p_{\alpha+}  -  \epsilon^{*}/\alpha I(\epsilon^{*}/\alpha > 0) \right)\exp\left\{ -\epsilon^{*} p_{\alpha+} + (\alpha p_{\alpha+})^{2}  \right\}   
\right) , 
\ee
Interestingly, for $\alpha = 0$, GAL is exactly the AL distribution (\cite{yan2017new}). Also noting that if $\epsilon \sim \text{GAL}(\alpha, \tau, \sigma)$, with $\epsilon \in \mathbf{R}$, then $\int^{0}_{-\infty} \text{GAL}(\epsilon| \mu, \alpha, \tau, \sigma) d\epsilon= \tau_{0} = h(\gamma)\tau \neq \tau$ where $h(\gamma) = 2\Phi(-|\gamma|)\exp\{0.5\gamma^2\}$ with $\gamma = (\bI(\alpha > 0) - \tau)|\alpha|$ and $\text{GAL}$ denote the density of the GAL distribution. This means that we will need to specify $\tau_0$ and then we get $\tau = \bI(\gamma < 0) + \{[\tau_0 - \bI(\gamma < 0)]/h(\gamma)\}$. We note that for the choice of $\tau_0 \in(0, 1)$, $\gamma$ is bounded and $\gamma \in (\gamma_L, \gamma_U)$, where $\gamma_L$ is the negative root of $h(\gamma) - (1-\tau_0)$ and $\gamma_U$ is the positive root of $h(\gamma) - \tau_0$, which can easily be obtained using any optimization routine in R like optim \citealp{yan2017new}. Finally, GAL has the following mixture representation as (so that the $\tau^{th}$ quantile is at zero) 
\be
\epsilon = C|\gamma| s + A\nu + u\sqrt{\sigma B\nu},&&\text{where}\; s \sim \Normal^{+}(0, \sigma^2),\; \nu \sim \text{Exp}(\sigma),\; u \sim \Normal(0, 1),
\ee
The use of GAL as a flexible distribution for quantile regression is growing. \cite{rahman2019flexible} consider GAL in the case of quantile regression modeling for ordinal data; \cite{kobayashi2021flexible} considered a Dirichlet process mixture distribution of GAL (denote the GAL Dirichlet mixture process as GALDP) in the Bayesian quantile regression. The mixture distribution is is over the parameter vector $(\gamma, \sigma) \in [\gamma_{L}, \gamma_{U}] \times \mathcal{R}^{+}$. We adopt similar truncated model formulation here. Namely, we assume $\epsilon_{i} \stackrel{iid}{\sim} \uG$, where 
\be
\epsilon_{i} \stackrel{iid}{\sim} f_{\epsilon}(\epsilon|\uOmega_{\epsilon})  &=& \sum^{K\epsilon}_{k =1} \pi_{\epsilon,k}\text{GAL}(\tau_0, \gamma_{k}, \sigma_k)
\ee
where $\uOmega_{\epsilon} = \{ (\pi_{\epsilon,k}, \gamma_{\epsilon, k}, \sigma_{\epsilon, k})\}^{K_{\epsilon}}_{k=1}$
for some chosen $K_{\epsilon} < \infty$. Using the truncated construction of \cite{sethuraman1994constructive}, we then have $( \pi_{\epsilon,1}, \cdots,  \pi_{\epsilon,K_n})\trans \sim \Dir(\alpha/K_n, \cdots, \alpha/K_n)$ and $(\gamma_{\epsilon, 1}, \sigma_{\epsilon, 1})$, $\cdots, (\gamma_{\epsilon, K_\epsilon}, \sigma_{\epsilon, K_{\epsilon}}) \stackrel{iid}{\sim} \Unif(\gamma_{L}, \gamma_{U})\Ga(a_0, b_0)$, with the concentration parameter $\alpha > 0$. 

\subsection{Priors Specifications}
We now discuss the choice of priors for all model parameters. For $\ubeta_{z} \sim \MVN(\bzero, \Sigma_{z,0})$; for the vector of parameters $\uphi = (\phi_1, \cdots, \phi_{K_n})\trans$, we use a Bayesian P-splines priors as proposed by \cite{lang2004bayesian} and assume $\uphi \sim \MVN(\bzero, \theta^{2}\uP^{-1})$, where $P$ is the second order difference matrix. Additionally, we use the approach proposed by \cite{klein2016scale} for the prior for $\theta^2$. 
For the measurement error vector, $ \uU_i \stackrel{iid}{\sim}\sum^{K_u}_{k=1} \pi_{u,k}\MVN(.|\umu_{u,k}, \Sigma_{k,u}), $ with $\umu_{u,k} \sim \MVN(.|\umu_{u,0}, \Sigma_{u,0})$, constrained to $\sum^{K_\epsilon}_{k=1}  \pi_{u,k} \umu_{u,k} = \bzero$;  $\Sigma_{k,u} \sim \mbox{Inv-W}(.|\nu_{u,0}, \Psi_{0,u})$; and $\upi_{u} = (\pi_1, \cdots, \pi_{K_u})\trans \sim \Dir(\alpha_u/K_u, \cdots, \alpha_u/K_u)$.

For the error model of the IV, $ \uomega^{*}_i \stackrel{iid}{\sim}  \sum^{K_\omega}_{k=1} \pi_{\omega,k}\MVN(.|\umu_{\omega,k}, \Sigma_{\omega,k})$, $\umu_{\omega,k} \sim \MVN(.|\umu_{\omega,0}, \Sigma_{\omega,0})$, and  $\Sigma_{\omega,k} \sim \mbox{Inv-W}(.|\nu_0, \Psi_{0,\omega})$ constrained to $\sum^{K_\omega}_{k=1}  \pi_{\omega,k} \umu_{\omega,k} = \bzero$. 

For the cluster probability, we assume $\upi_{\omega} = (\pi_1, \cdots, \pi_{K_\omega})\trans \sim \Dir(\alpha_u/K_\omega, \cdots, \alpha_u/K_\omega)$. Additionally, we assume $\uX_i \sim  \sum^{K_x}_{k=1} \pi_{x,k} \MVN(.|\umu_{x,k}, \Sigma_{x,k})$, with $\umu_k \sim \MVN(.|\umu_0, \Sigma_0)$, $\Sigma_{x,k} \sim \mbox{Inv-W}(.|\nu_{x,0}, \Psi_0)$, where
$\upi_{x} = (\pi_1, \cdots, \pi_{K_x})\trans \sim \Dir(\alpha_x/K_x, \cdots, \alpha_x/K_x)$.

\subsection{Joint posterior distribution}
We first define the following parameters. Let's $\uOmega_{u} = \{ (\pi_{u,k}, \umu_{u, k}, \Sigma_{u, k})\}^{K_{u}}_{k=1}$, $\uOmega_{\omega} = \{ (\pi_{\omega,k}, \sigma_{\omega, k})\}^{K_{\omega}}_{k=1}$, $\uOmega_{x} = \{ (\pi_{\omega,k}, \sigma_{x, k})\}^{K_{x}}_{k=1}$, and $\uOmega_{\epsilon}$ as defined above. The joint likelihood of the data conditional on all the parameters is proportional to
\be
\ell(\uY_i, \uW_i, \uM_i|\tau_0, \nu_i, s_i, \uX_i, \uOmega_{\epsilon},\uOmega_{u},\uOmega_{\omega},\uOmega_{x})&\propto& f_{GAL}(Y_{i}|\tau_0, \nu_i, s_i, \uX_i, \uOmega_{\epsilon}) \nonumber \\
&& f_{u_i}(\uW_i|\uX_i, \uOmega_{u})f_{\omega_i}(\uM_i|\uX_i, \uOmega_{\omega}). \nonumber
\ee
Based on that likelihood, the joint posterior distribution is obtained as proportional to 
\be
Pr(\unu, \us, \uX, \uOmega_{\epsilon},\uOmega_{u},\uOmega_{\omega},\uOmega_{x}| Data) &\propto& \prod^{n}_{i=1}\left\{ \ell(\uY_i, \uW_i, \uM_i|\tau_0, \nu_i, s_i, \uX_i, \uOmega_{\epsilon},\uOmega_{u},\uOmega_{\omega},\uOmega_{x}) \right. \nonumber \\
&&\times\left. Pr(\uX_i|\uOmega_{x}) Pr(\nu_i, s_i)\right\} Pr(\beta_0, \ubeta_z) \\
&&\times Pr(\uOmega_{\epsilon}) Pr(\uOmega_{u}) Pr(\uOmega_{\omega}) Pr(\uOmega_{x}) \nonumber
\ee
Directly sampling from the joint posterior is complicated and we sample from the joint posterior distribution using a sequence of Gibbs steps. We adopt the Metropolis-Hasting(MH) withing partially collapsed Gibbs samples approach considered in \cite{rahman2019flexible} to update the parameters $(\gamma_k, \sigma_k)$. 

\section{Simulation and Results} \label{sec:simul}
\subsection{Simulation Set-up}
For each $i =1,\cdots, n$ independently, we simulated $X_i(t)$ from a Gaussian process with $E\{X_i(t)\} = \{\sin(2\pi t) + 1.25\}/2$, $\var\{X_i(t)\} = \sigma^2_x $, and $\cov \{ X_i(t_l), X_i(t_j) \} = \rho_x \sigma^2_x $. We simulated $W_{ij}(t) = X_i(t) + U_{ij}(t)$ with $j = 1,\cdots,J = 5$, where $U(t)$ is a Gaussian process with $E\{U(t)\} = 0$, $\var\{U_i(t)\} = \sigma^2_u $, and $\cov\{U_i(t_l), U(t_j)\} = \rho_u \sigma^2_u$; 
Finally, we simulated the response for each unit as $Y_i = \uZ_{i}\trans\ubeta_z + \int^1_{0} \ubeta(t)X_{i}(t)dt + \epsilon_i$, where $\epsilon_i \sim \uG$, for some distribution $\uG$. 
In each simulation exercise, we assessed the performance of each estimator of $\ubeta_{\tau}(t)$ ($\tau = 0.25, 0.50, 0.90$) with the average bias squared ($\text{ABias}^2$), average sample variance (\text{Avar}), and the  mean squared integrated error (\text{MISE}):
\be
ABias^2 (\widehat{\beta}_{\tau}) &=& \frac{1}{n_{grid}}\sum^{n_{grid}}_{l=1}\left\{ \bar{\beta}_{\tau}(t_l) - \beta_{\tau}(t_l) \right\}^2, \nonumber\\
Avar(\widehat{\beta}_{\tau}) &=& \frac{1}{n_{s} n_{grid} }\sum^{n_{r} }_{i=1}\sum^{n_{grid}}_{j=1}\left\{ \widehat{\beta}_{\tau}(t_j) - \bar{\beta}_{\tau}(t_j)\right\}^2 {, and } \nonumber \\
MISE &=& ABias^2 + Avar, \nonumber
\ee
where $n_{grid}$ represents the number of equally selected grid points between 0 and 1 and $\bar{\beta}_{\tau}(t_j) = \frac{1}{n_r}\sum^{n_{r}}_{j=1}\widehat{\beta}_{\tau}^{r}(t_j)$ represents the point-wise average over the $n_r$ replicates of $\beta_{\tau}(t_j)$ at a specific time point $t_j$.
We denote the estimate of $\beta_{\tau}(t)$ from our approach as $\widehat{\bbeta}_{\tau, fBQ}(t)$ (full Bayesian approach). We also consider the naive approach which treats $\uW$ as a precisely observed covariate and uses $\uW$ as the true covariate $\uX$. The estimates obtained from the naive approach will be denoted as $\widehat{\bbeta}_{\tau, naive}(t)$. One potential drawback of our fully Bayesian estimation approach is the speed, especially in the case of large dataset. Another approach we consider here a two stage approach. In the first stage, we approximate the value of functional covariate $\uX$ based on its observed proxy data $\uW$ and using a mixed effect model. Then in the second stage, estimation is done using a Bayesian approach only using model~\ref{eq4} . More details on this approach is provided in \cite{cui2022fast}. We will denote parameter estimates obtained using this approach as $\widehat{\bbeta}_{\tau, fast}(t)$.  For all of our simulation, we use the mixture of three GAL component.  

\subsection{Simulation Results}
Our simulation exercise investigate various aspects of our proposed approach and is divided into different cases. For all cases, unless otherwise specified, we assume $\sigma_{x} = 4$, $\rho_x = 0.5$, $\sigma_u = 4$, $\rho_u = 0.5$. 
In simulation {\bf Case 1}, we consider a simple case and assume $\uG$ (the error distribution for the response Y) is the normal distribution with mean zero and standard deviation $\sigma_e = 1$. We consider the following sample sizes n = 200, 500, and 1000. We report the performance of the approaches considered in Table~\ref{Tab:tab1}.  Across the quantiles considered, the full Bayesian approach, $\widehat{\bbeta}_{\tau, fBQ}(t)$ and the RC approach, $\widehat{\bbeta}_{\tau, fast}(t)$, perform very similarly; although the fully Bayesian estimator tends to have higher mean squared error (MISE) for smaller sample sizes but smaller MISE larger sample sizes. The estimator based on the naive approach, $\widehat{\bbeta}_{\tau, naive}(t)$, performs the best for smaller sample sizes but has the highest MISE for larger sample sizes. Suggesting a lost in efficiency when using the naive approach when compared to a measurement error adjusted approach. As expected the bias tended to high for the naive approach and remain high even for large sample sizes.
\begin{table}[ht]
\centering
\begin{tabular}{|r|lrrrr|}
  \hline
 & Method & n &  bias$^{2}$ & AVar & MISE \\ 
  \hline
\multirow{9}{*}{$\tau$ = 0.25} & \multirow{3}{*}{$\beta_{\tau, fast}(t)$} & 200 &   0.691 & 0.023 & 0.714 \\ 
   &  & 500 &   0.048 & 0.003 & 0.051 \\ 
   &  & 1000 &   0.057 & 0.001 & 0.059 \\ \cline{2-6}
  &  \multirow{3}{*}{$\beta_{\tau, naive}(t)$} & 200 &   0.688 & 0.012 & 0.700 \\ 
  &  & 500 &   0.134 & 0.002 & 0.136 \\ 
  &  & 1000 &   0.127 & 0.001 & 0.128 \\ \cline{2-6}
  & \multirow{3}{*}{$\beta_{\tau, FBQ}(t)$} & 200 &   0.969 & 0.042 & 1.011 \\ 
   &  & 500 &   0.036 & 0.008 & 0.044 \\ 
   &  & 1000 &   0.032 & 0.003 & 0.035 \\  \hline 
  \multirow{9}{*}{$\tau$ = 0.50} & \multirow{3}{*}{$\beta_{\tau, fast}(t)$} & 200 &   0.693 & 0.018 & 0.711 \\ 
   &  & 500 &   0.071 & 0.002 & 0.074 \\ 
   &  & 1000 &   0.078 & 0.001 & 0.079 \\ \cline{2-6}
   &  \multirow{3}{*}{$\beta_{\tau, naive}(t)$} & 200 &   0.708 & 0.010 & 0.718 \\ 
   &  & 500 &   0.176 & 0.002 & 0.178 \\ 
   &  & 1000 &   0.167 & 0.001 & 0.168 \\ \cline{2-6}
 & \multirow{3}{*}{$\beta_{\tau, FBQ}(t)$} & 200 &   0.869 & 0.036 & 0.905 \\ 
   &  & 500 &   0.044 & 0.007 & 0.050 \\ 
   &   & 1000 &   0.038 & 0.003 & 0.041 \\  \hline
   \multirow{9}{*}{$\tau$ = 0.90} & \multirow{3}{*}{$\beta_{\tau, fast}(t)$} & 200 &   0.600 & 0.036 & 0.636 \\ 
   &   & 500 &   0.039 & 0.005 & 0.044 \\ 
   &   & 1000 &  0.043 & 0.003 & 0.046 \\ \cline{2-6} 
  &  \multirow{3}{*}{$\beta_{\tau, naive}(t)$} & 200 &   0.663 & 0.019 & 0.682 \\ 
   &   & 500 &   0.104 & 0.003 & 0.107 \\ 
   &  & 1000 &  0.091 & 0.002 & 0.093 \\  \cline{2-6}
   & \multirow{3}{*}{$\beta_{\tau, FBQ}(t)$} & 200 &   1.391 & 0.083 & 1.474 \\ 
   &   & 500 &   0.055 & 0.015 & 0.070 \\ 
   &  & 1000 &   0.070 & 0.006 & 0.076 \\ 
   \hline
\end{tabular}
\caption{Simulation summaries for {\bf Case 1} with varying sample sizes. $\beta_{\tau, FBQ}(t)$ denote our fully Bayesian approach; $\beta_{\tau, fast}(t)$ the fast Regression calibration approach; $\beta_{\tau, naive}(t)$ the naive approach which ignores the measurement error.} 
\label{Tab:tab1}
\end{table}

In {\bf Case 2} of our simulation, we consider the case of a skewed distribution for the error. We first assume that the error terms are from a skew t-distribution with location parameter $\xi = 0$, degrees-of-freedom $\nu = 5$, and slant parameter $\alpha = 2.0$. The skew t-distribution can easily be simulated in R using the package \cite{snpackage}. Assuming again the same setting as in {\bf Case 1} and sample size n = 500. The results of {\bf Case 2} scenarios are shown in Table~\ref{tab2}. In summary, the full Bayesian and regression calibration (RC)  approach performs very similarly, although the full Bayesian approach perform best. Again the naive approach performs the worse with a high bias.   

\begin{table}[ht]
\centering
\begin{tabular}{|rlrlrrr|} \hline
 & &  & bias$^{2}$ & Var & MISE \\ \hline
& \multirow{3}{*}{$\tau = 0.25$}& $\beta_{\tau, fast}(t)$ & 0.049 & 0.001 & 0.050 \\ 
  &  &$\beta_{\tau, naive}(t)$ & 0.130 & 0.001 & 0.131 \\ 
   & & $\beta_{\tau, FBQ}(t)$ & 0.030 & 0.003 & 0.033 \\ \hline
& \multirow{3}{*}{$\tau = 0.50$} & $\beta_{\tau, fast}(t)$ & 0.054 & 0.001 & 0.055 \\ 
   & & $\beta_{\tau, naive}(t)$ & 0.146 & 0.001 & 0.147 \\ 
   & & $\beta_{\tau, FBQ}(t)$ & 0.034 & 0.002 & 0.036 \\  \hline
 &\multirow{3}{*}{$\tau = 0.90$} & $\beta_{\tau, fast}(t)$ & 0.043 & 0.002 & 0.044 \\ 
  & & $\beta_{\tau, naive}(t)$ & 0.110 & 0.001 & 0.111 \\ 
   & & $\beta_{\tau, FBQ}(t)$ & 0.031 & 0.003 & 0.034 \\  \hline 
\end{tabular}
\caption{Summary of simulation for {\bf case 2}. The data were simulated under the skew t-distribution with location parameter $\xi = 0$, degrees-of-freedom $\nu = 5$, and slant parameter $\alpha = 2.0$. Simulation were done assuming $n = 500$.} 
\label{tab2}
\end{table}

In {\bf Case 3}, we investigate the impact of moderate to large measurement error variance $\sigma_u$ on the functional parameter estimate $\beta_{\tau}(t)$. We run the simulation assuming values of $\sigma_u = 1, 4$, and $16$. We report the results in Table~\ref{tab3}. The performance of the naive estimator deteriorate with increasing measurement error (increasing MSIE), suggesting the need to adjust for measurement error. We note that although our fast measurement error adjusted approach tended to have high MSE in the case of low measurement error variance ($\sigma_u = 1$) when compared to the naive approach, the full Bayesian approach, however, performed very similarly than the naive approach for smaller measurement error but outperforms the naive approach for large ME variances. For larger ME variance $\sigma_u$, the naive approach tended to perform poorly (large $MISE$)  

\begin{table}[ht]
\centering
\begin{tabular}{|rrlrrr|}
  \hline
   &  &    & $Bias^{2}$ & Var & MISE \\ 
  \hline
  \multirow{9}{*}{$\sigma_u = 1.0$} & \multirow{3}{*}{$\tau = 0.25$}& $\beta_{\tau, fast}(t)$ & 0.065 & 0.002 & 0.068 \\ 
     & & $\beta_{\tau, naive}(t)$ & 0.034 & 0.002 & 0.036 \\ 
     &   & $\beta_{\tau, FBQ}(t)$ & 0.050 & 0.005 & 0.056 \\ \cline{3-6}
      & \multirow{3}{*}{$\tau = 0.5$} & $\beta_{\tau, fast}(t)$ & 0.072 & 0.002 & 0.074 \\ 
     &    & $\beta_{\tau, naive}(t)$ & 0.036 & 0.001 & 0.038 \\ 
      &    & $\beta_{\tau, FBQ}(t)$ & 0.039 & 0.004 & 0.044 \\ \cline{3-6}
      & \multirow{3}{*}{$\tau = 0.9$} & $\beta_{\tau, fast}(t)$ & 0.045 & 0.004 & 0.049 \\ 
     &    & $\beta_{\tau, naive}(t)$ & 0.035 & 0.004 & 0.039 \\ 
      &    & $\beta_{\tau, FBQ}(t)$ & 0.095 & 0.009 & 0.104 \\ \hline \hline
     \multirow{9}{*}{$\sigma_u = 4.0$} & \multirow{3}{*}{$\tau = 0.25$}& $\beta_{\tau, fast}(t)$ & 0.048 & 0.003 & 0.051 \\ 
      &   &   $\beta_{\tau, naive}(t)$ & 0.134 & 0.002 & 0.136 \\ 
      &   &   $\beta_{\tau, FBQ}(t)$ & 0.036 & 0.008 & 0.044 \\ \cline{3-6}
      & \multirow{3}{*}{$\tau = 0.50$} &   $\beta_{\tau, fast}(t)$ & 0.071 & 0.002 & 0.074 \\ 
      &   &   $\beta_{\tau, naive}(t)$ & 0.176 & 0.002 & 0.178 \\ 
      &   &   $\beta_{\tau, FBQ}(t)$ & 0.044 & 0.007 & 0.050 \\ \cline{3-6}
      & \multirow{3}{*}{$\tau = 0.90$} & $\beta_{\tau, fast}(t)$ & 0.039 & 0.005 & 0.044 \\ 
      &   &    $\beta_{\tau, naive}(t)$ & 0.104 & 0.003 & 0.107 \\ 
      &   &    $\beta_{\tau, FBQ}(t)$ & 0.055 & 0.015 & 0.070 \\  \hline \hline
     \multirow{9}{*}{$\sigma_u = 16.0$} & \multirow{3}{*}{$\tau = 0.25$}& $\beta_{\tau, fast}(t)$ & 0.480 & 0.009 & 0.489 \\ 
      &   &   $\beta_{\tau, naive}(t)$ & 0.830 & 0.001 & 0.831 \\ 
      &   &   $\beta_{\tau, FBQ}(t)$ & 0.733 & 0.013 & 0.747 \\ \cline{3-6}
     & \multirow{3}{*}{$\tau = 0.50$} &   $\beta_{\tau, fast}(t)$ & 0.703 & 0.006 & 0.709 \\ 
      &   &  $\beta_{\tau, naive}(t)$ & 0.900 & 0.000 & 0.901 \\ 
     &   &   $\beta_{\tau, FBQ}(t)$ & 0.867 & 0.007 & 0.874 \\ \cline{3-6}
      & \multirow{3}{*}{$\tau = 0.90$} &  $\beta_{\tau, fast}(t)$ & 0.270 & 0.020 & 0.290 \\ 
      &   &   $\beta_{\tau, naive}(t)$ & 0.747 & 0.001 & 0.748 \\ 
      &   &   $\beta_{\tau, FBQ}(t)$ & 0.436 & 0.052 & 0.488 \\ 
   \hline
\end{tabular}
\caption{ Summary of simulation for {\bf case 3}. The data are simulated assuming varying level of measurement error variance ($\sigma_u = 1, 4, 16$). } 
\label{tab3}
\end{table}

Finally in {\bf Case 4}, we study the impact of the number of replicates on the performance of the estimators considered since both ME corrected approaches(full Bayesian and the fast univariate) require at least two replicates.
We perform the simulation similar to {\bf case 1} but set $n=500$ and we allow the number of replicate for individual to vary from $J = 2, 3, 4$. In Table~\ref{tab:tab4}, overall both ME corrected approach perform very similarly for larger number of replicates when compared to lower number of replicates. Again both ME corrected approach outperform the naive approach in all most settings. 
\begin{table}[ht]
\centering
\begin{tabular}{|rllrrrrr|}
  \hline
 &   &   & Methods &  Bias$^2$ & AVar & MISE \\ 
  \hline
  & \multirow{9}{*}{$J = 2$} &  \multirow{3}{*}{$\tau = 0.25$}& $\beta_{\tau, fast}(t)$ & 0.073 & 0.004 & 0.077 \\ 
    &   & & $\beta_{\tau, naive}(t)$ &   0.307 & 0.002 & 0.309 \\ 
    &   &  & $\beta_{\tau, FBQ}(t)$ &   0.162 & 0.036 & 0.197 \\ \cline{4-7}
    &   & \multirow{3}{*}{$\tau = 0.25$} & $\beta_{\tau, fast}(t)$ &   0.145 & 0.003 & 0.148 \\ 
    &   &   & $\beta_{\tau, naive}(t)$   & 0.380 & 0.001 & 0.382 \\ 
    &   &   & $\beta_{\tau, FBQ}(t)$   & 0.230 & 0.029 & 0.259 \\ \cline{4-7}
    &   & \multirow{3}{*}{$\tau = 0.90$} & $\beta_{\tau, fast}(t)$   & 0.039 & 0.006 & 0.045 \\ 
    &  &   & $\beta_{\tau, naive}(t)$   & 0.237 & 0.003 & 0.240 \\ 
    &   &   & $\beta_{\tau, FBQ}(t)$   & 0.282 & 0.063 & 0.344 \\  \hline
    & \multirow{9}{*}{$J = 3$} & \multirow{3}{*}{$\tau = 0.90$} & $\beta_{\tau, fast}(t)$ &   0.053 & 0.003 & 0.057 \\ 
    &   &   & $\beta_{\tau, naive}(t)$   & 0.214 & 0.002 & 0.216 \\ 
    &   &   & $\beta_{\tau, FBQ}(t)$&   0.060 & 0.023 & 0.083 \\ \cline{4-7}
    &   & \multirow{3}{*}{$\tau = 0.50$} & $\beta_{\tau, fast}(t)$ &   0.097 & 0.003 & 0.100 \\ 
    &   &   & $\beta_{\tau, naive}(t)$ &   0.272 & 0.001 & 0.273 \\ 
    &   &   & $\beta_{\tau, FBQ}(t)$ &   0.087 & 0.020 & 0.107 \\ \cline{4-7}
    &   & \multirow{3}{*}{$\tau = 0.90$} & $\beta_{\tau, fast}(t)$ &   0.036 & 0.006 & 0.042 \\ 
    &   &   & $\beta_{\tau, naive}(t)$   & 0.165 & 0.003 & 0.168 \\ 
    &   &   & $\beta_{\tau, FBQ}(t)$ &  0.122 & 0.044 & 0.166 \\ \hline
    & \multirow{9}{*}{$J = 4$} & \multirow{3}{*}{$\tau = 0.25$} & $\beta_{\tau, fast}(t)$ &   0.048 & 0.003 & 0.051 \\ 
    &   &   & $\beta_{\tau, naive}(t)$ &   0.161 & 0.002 & 0.162 \\ 
    &   &   &$\beta_{\tau, FBQ}(t)$   & 0.041 & 0.012 & 0.053 \\ \cline{4-7}
    &   & \multirow{3}{*}{$\tau = 0.50$} & $\beta_{\tau, fast}(t)$ &   0.078 & 0.002 & 0.080 \\ 
   &   &   & $\beta_{\tau, naive}(t)$ &   0.212 & 0.002 & 0.213 \\ 
    &   &   & $\beta_{\tau, FBQ}(t)$ &   0.051 & 0.011 & 0.062 \\ \cline{4-7}
    &   & \multirow{3}{*}{$\tau = 0.90$} & $\beta_{\tau, fast}(t)$ &   0.034 & 0.005 & 0.040 \\ 
    &   &   & $\beta_{\tau, naive}(t)$ &   0.123 & 0.003 & 0.126 \\ 
    &   &   & $\beta_{\tau, FBQ}(t)$ &   0.076 & 0.024 & 0.100 \\ 
   \hline
\end{tabular}
\caption{ Summary of simulation for {\bf Case 4}. The data are simulated assuming the settings under {\bf Case 1} with $n=500$ and varying number of replicates $J$ ($J = 2, 3, 4$). } 
\label{tab:tab4}
\end{table}

\section{Application}\label{sec:application}
We revisit our case study where we had for goal to estimate the effect of average physical activity pattern (behavior) at various BMI quantile of US adults over 50 years in the NHANSES data set.  
Since the 2012, NHANES has started to include more objective (device based) measures of physical activity data in addition to self-reported PA behaviors. Study participants are asked to wear a PA monitor device for 24hrs on the non-dominant wrist for at least a week. Participants are then asked to return the PA monitor devices. The data are subsequently extracted and then process according to NHANES data quality checks \cite{nhanse22}. Four our analysis, we used the Monitor-Independent Movement Summary (MIMS) triaxial summary PA reported per minutes for each individuals. For complete data, this results in $1440$ observations per individual per day. NHANES also report the data quality flags and the predicted physical activity status for each minutes (Wake/sleep/non-wear) prediction. Following the data processing steps outlined in \cite{karas2019accelerometry}, we only consider individuals data with no quality flags, individuals with at least 3 valid week days (Monday to Friday). A valid day is defined as a day with no more that 10\%(144 minutes) of invalid minutes. A minute is designated as valid if it has triggered zero quality flags and is classified as wear. To remove the effect of extreme values on the analysis, the data is winsorized and MIMS values above the $99.9$ percentile of the data are simply replaced by the $99.9$ percentile. Finally, minutes identified as non-wear or with value $-0.01$ are labelled as missing values. Similar to  \cite{karas2019accelerometry,cui2022semiparametric}, we use the function \textit{fpca.ssvd} in the R package Refund in R \cite{Rcomp} to obtain the predicted the PA for each individual. Subsequently, missing values are replaced by their predicted values obtained from \textit{fpca.sc}.  

We fit our Bayesian quantile regression model to estimate the $0.1$, $0.5$, and $0.9$ quantiles of BMI as a function of the following error free covariates: Gender (Male =Reference/Female = 0); Race(Black =Reference/Hispanic/White/Mic/Other), Self reported Health condition (Excellent = Reference / Good / fair and poor); Age (in years). We would also like to adjust for usual weekdays physical activity pattern on BMI, which is unfortunately unobserved, but can be approximated by weekdays physical activity pattern. We first correct for measurement error in the observed weekdays physical activity pattern $\uW_{ij}(t)$ using the fast version using the approach proposed by \cite{cui2022fast}. We then get an estimate of $\widehat{X}_{i}(t)$, which are then used as functional covariate input in our Bayesian quantile regression model. As a second option, we fit the similar Bayesian quantile regression model simply averaging the physical activity pattern for each individual across the 5 week days. Subsequently, these $\bar{W}_{i}(t)$ as then used as input for a naive Bayesian quantile regression model. 
We fit multiple Bayesian model with varying number of GAL mixtures including a model with only one GAL component. The model with one GAL component tended to fit these data best, based on the WAIC.  We perform model check using posterior predictive checks as suggested in the Bayesian workflow \citep{gelman2020bayesian}.  The model check reveal no issues.  For easy implementation of our approach, we have written a R code implementing our approach in the widely use brm function in the Bayesian analysis package Brms \citep{burkner2017brms} and the Bayesian analysis package Rstan \citep{carpenter2017stan}.  

\begin{figure}[!ht]
    \centering
    \includegraphics[scale=.5]{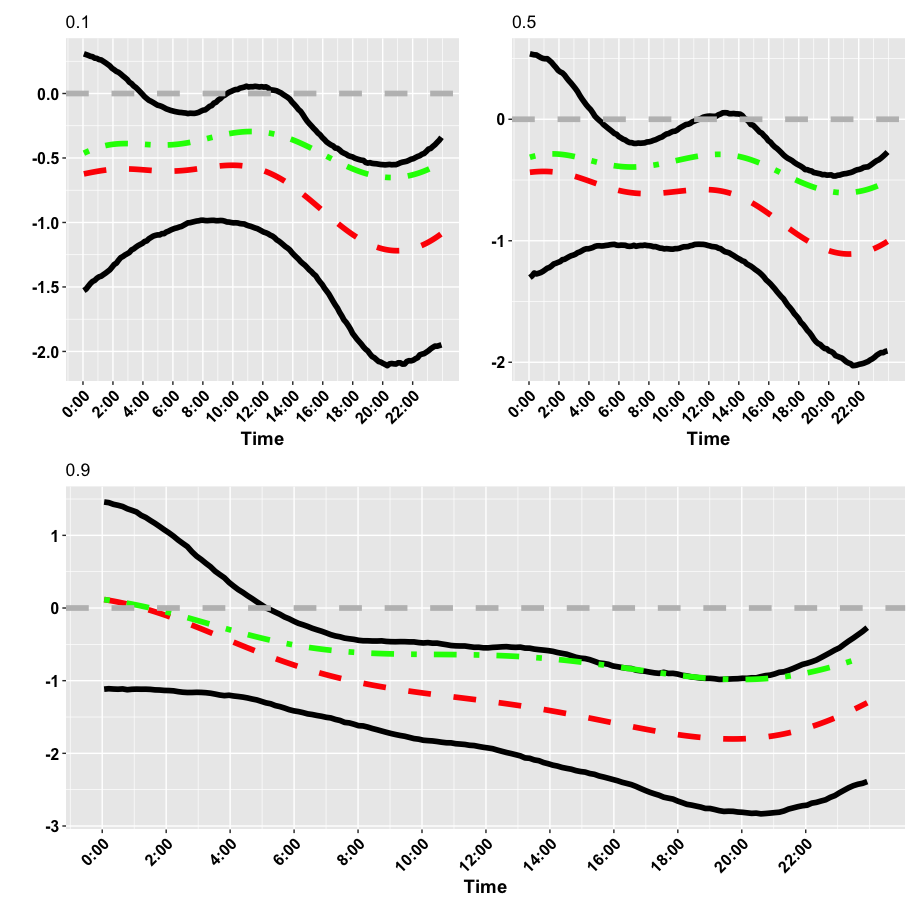}
    \caption{Plot of the posterior estimates of $\bbeta(t)$. Dash (red) denote the estimate obtained after the measurement correction (RC - ). Dotted (green) denote the estimate obtained under the naive approach (no measurement error adjustment). Continue line (black) represent the bound of the point-wise 95\% posterior credible intervals.}
    \label{fig:finalbetat}
\end{figure}

We report the plot of the estimate $\bbeta(t)$ with and without measurement error correction on Figure~\ref{fig:finalbetat}. We observed that all quantiles, the estimated weight $\bbeta(t)$ are all negative for most part. However, we note that measurement error corrected estimate lies below all the naive approach for all quantiles. In fact the weights from the naive seem attenuated towards zero for all quantiles considered when compared to the measurement error corrected approach. This is characteristic of the attenuation effect observed in covariates measured with errors. This potentially shows the importance of adjusting for measurement error in these device measured physical activity reported as functional covariates. 
Additionally, we also reported the posterior inference for the error free covariates (see Table~\ref{tab:finalSumT}).

\begin{table}[ht]
\centering
\resizebox{\textwidth}{!}{\begin{tabular}{l|rr|rr|rr|}
  \hline
    &\multicolumn{2}{|c|}{$\tau = 0.1$}& \multicolumn{2}{|c|}{$\tau = 0.5$}& \multicolumn{2}{|c|}{$\tau = 0.9$} \\ \hline
      & mean &(2.5\% - 97.5\%) & mean &(2.5\%-97.5\%)& mean&(2.5\%-97.5\%) \\ \hline
    Intercept & 34.30 & (31.31 , 37.19) & 39.83 & (36.92 , 42.67) & 58.01 & (54.28, 61.62) \\ 
    Gender (Male) & -0.50 & (-1.07 , 0.06) & -0.71 & (-1.28,  -0.13)& -2.25 & (-2.97 , -1.53) \\ 
    Race - Baseline (Black) & - & - & - & - & - & - \\
    Race (Hispanic) & 0.05 & (-0.86 , 0.95) & -0.26 & (-1.18, 0.64) & -1.99 & (-3.16 , -0.82) \\ 
    Race (Mix/Other) & -3.01 & (-4.09 , -1.93) & -3.27 & (-4.34 , -2.26) & -5.79 & (-7.06 , -4.52) \\ 
    Race (White) & -0.58 & (-1.32 , 0.17) & -0.80 & (-1.58 , -0.03) & -1.59 & (-2.58 , -0.62) \\ 
    Health Condition - Baseline (Excellent) & - & - & - & - & - & - \\
    Health Condition (Fair or Poor) & 0.83 & (0.05 , 1.64) & 0.78 &( 0.01,  1.57) & 2.39 & (1.41 , 3.38) \\ 
    Health Condition (Good) & 0.78 & (0.13 , 1.43) & 0.69 & (0.06 , 1.34) & 0.81 & (0.01 , 1.63) \\ 
    Age(Year) & -0.11 & (-0.14 , -0.075) & -0.11 &(-0.14 , -0.075) & -0.21 & (-0.25, -0.17) \\ 
   \hline
\end{tabular}}
\caption{Posterior mean along with the 95\% posterior credible intervals respectively for the quantiles $\tau =0.1, 0.5, 0.9$ of BMI based on the fast univariate RC measurement error correction approach.}
\label{tab:finalSumT}
\end{table}

\section{Conclusion} \label{sec:concl}
We propose a Bayesian measurement error corrected approach in a quantile regression setting when the functional covariates are measured with errors based on two approaches. One approach is based on the full model and the other is based on a regression calibration approach. Through a set of simulation settings, we show the importance to adjust/correct for measurement errors in the functional covariates. Averaging across the replicated functions (naive approach) does not at all remove the measurement error and result in a biased estimate of the functional coefficient. In our application, we show that our measurement error corrected lead to a less attenuated functional effect when we apply our ME corrected approach compared to using the naive approach. Although our full Bayesian ME correction approach seem very suitable since it reflects all model uncertainty, it does not scale with increasing sample size. For this reason we recommend using our fast univariate approach. For easy implementation of our approach, we have will share the R code used in the analysis on the author's Github.   

\clearpage
\newpage

 \bibliographystyle{plainnat}
\bibliography{RogerReference,BQSOFReg}
\end{document}